\documentclass[pra,onecolumn,preprint,showpacs,groupedaddress,superscriptaddress,nofootinbib,floatfix,preprintnumbers,longbibliography,aps]{revtex4-2}

\usepackage{physics}
\usepackage{xcolor}
\usepackage{graphicx}% Include figure files
\usepackage{subfigure} % subfiguras
\usepackage{dcolumn}% Align table columns on decimal point
\usepackage{bm}% bold math

\usepackage{comment}

\begin{document}

\title{Orbital Angular Momentum Beam assisted High-Order Harmonic Generation in Semiconductor Materials}

\author{C. Granados}
\affiliation{Department of Physics, Guangdong Technion - Israel Institute of Technology, 241 Daxue Road, Shantou, Guangdong, China, 515063}
\email{camilo.granados@gtiit.edu.cn}
\affiliation{Technion - Israel Institute of Technology, Haifa, 32000, Israel}
\affiliation{Guangdong Provincial Key Laboratory of Materials and Technologies for Energy Conversion, Guangdong Technion - Israel Institute of Technology, 241 Daxue Road, Shantou, Guangdong, China, 515063}

\author{B. Kumar Das}
\affiliation{Department of Physics, Guangdong Technion - Israel Institute of Technology, 241 Daxue Road, Shantou, Guangdong, China, 515063}
\affiliation{Technion - Israel Institute of Technology, Haifa, 32000, Israel}
% \affiliation{Department of Physics and Solid State Institute, Technion -- Israel Institute of Technology, Haifa, 32000, Israel}
\affiliation{Guangdong Provincial Key Laboratory of Materials and Technologies for Energy Conversion, Guangdong Technion - Israel Institute of Technology, 241 Daxue Road, Shantou, Guangdong, China, 515063}

\author{M. F. Ciappina}
\email{marcelo.ciappina@gtiit.edu.cn}
\affiliation{Department of Physics, Guangdong Technion - Israel Institute of Technology, 241 Daxue Road, Shantou, Guangdong, China, 515063}
\affiliation{Technion - Israel Institute of Technology, Haifa, 32000, Israel}
\affiliation{Guangdong Provincial Key Laboratory of Materials and Technologies for Energy Conversion, Guangdong Technion - Israel Institute of Technology, 241 Daxue Road, Shantou, Guangdong, China, 515063}

\begin{abstract}

We investigate the use of light beams carrying orbital angular momentum (OAM) in the context of high harmonic generation (HHG) within semiconductor crystals. Our contribution deals with the transfer and conservation of OAM in the strong-field regime, from the driving laser field to the generated harmonics. To this end, in this work, we combine the semiconductor Bloch equations with the thin slab model to simulate the generation of high-order harmonics in semiconductor media and to compute the features of the far-field harmonics. We demonstrate that this theoretical approach is capable of satisfactorily reproducing previously published experimental features of the generated harmonics in ZnO driven by a Laguerre-Gauss beam. Our research not only deepens the understanding of light-solid interactions but also heralds the dawn of bright, structured XUV coherent radiation sources with unparalleled potential across diverse technological areas, paving the way for enhanced functionalities in fields such as microscopy, spectroscopy, and optical communication.
%he simulations were made through the time-dependent Schr\"odinger equation in the single-active-electron approximation complemented by a wavelet analysis. 

\end{abstract}

%\keywords{Suggested keywords}%Use showkeys class option if 
\maketitle
\clearpage
\section{Introduction}

High harmonic generation (HHG) has emerged as a pivotal phenomenon in the field of nonlinear optics, with significant implications spanning both basic scientific research and industrial applications. Initially explored in dilute atomic gases, HHG has recently garnered considerable attention in solid-state media, particularly in bulk semiconductor crystals \cite{Vampa14,Vampa15,Vampa17,Yue22,Trung16}. The distinctive characteristics of solid-state HHG, including its efficiency and potential for integration into high-repetition-rate nanoscale devices, have sparked intense investigation into its underlying mechanisms and practical implementations \cite{Gauthier19,Vampa15,Vampa17}.

Since its first observation in bulk semiconductor crystals in 2011 \cite{Ghimire2011}, solid-state HHG has been a subject of extensive research endeavors aimed at unraveling its fundamental principles, more recently devoted to the role of symmetries in the harmonic selection rules \cite{Oren23}. Unlike HHG in gases, the harmonics in solids are generated within a high-density, periodic crystal structure (in the last layers with a thickness of tens of nanometers). Additionally, solid-state HHG exhibits a lower intensity threshold compared to gas-phase HHG, making it particularly suitable for high-repetition-rate applications.

Due to the complexity of the interaction between the solid structure and the strong laser field, distinct theoretical models have been developed to comprehend the process effectively. The more renowned theories are the Time-dependent Schr\"odinger equation (TDSE) \cite{TDSE1,TDSE2,TDSE3}, the semiconductor Bloch equations (SBE) \cite{SBE1,SBE2,SBE3,Haug09}, and the time-dependent density functional theory (TDDFT) \cite{Runge84, TDDFT1,TDDFT2}. Both the TDSE and SBE models solve the laser-solid interaction problem under the single active electron (SAE) approximation. The TDDFT method offers a unique possibility of solving the many-body problem considering both electron-electron correlations and the full electronic band structure \cite{Correlations,Leeuwen99}. {Contrary to HHG in gases, where the strong field approximation (SFA) is widely used for HHG calculations, there is no a widely used theory for the calculations of solid HHG. The counterpart to the SFA for solid targets was proposed in Ref. \cite{Vampa14}. 
Hence, it is imperative to delve into new scenarios where our understanding of this intriguing phenomenon can be tested to converge to a single theoretical description.” 
}

The previously mentioned models enable us to address the microscopic aspects of solid-state HHG, specifically how individual emitters (electrons) respond to the strong laser field. However, there is an additional dimension to consider: the propagation of these generated high-order harmonics. It is well known that both gas-phase and solid-state HHG are collective phenomena. To obtain a measurable number of photons, the coherent sum and phase-matched emission of many emitters, typically ranging from $\sim10^{10}$ to $10^{12}$, are required~\cite{ddahhg}. Therefore, modeling both aspects presents a formidable computational challenge~\cite{Gaarde_2008}. To reduce the problem complexity, one could assume, for instance, that the phase-matching conditions are achieved and study the behavior of the near and far harmonic fields.  Additionally, in solid-state HHG, it is reasonable to assume that the high-order harmonic radiation is produced in the outermost layers of the material \cite{Ghimire3}, simplifying the modeling of its propagation even further.

In addition, one intriguing aspect of solid-state HHG lies in its capacity for tailored generation media, offering a unique degree of freedom absent in gas-phase HHG \cite{Gauthier19, You18}. Through precision patterning of the crystal surface, researchers can exert macroscopic control over the generation and emission of high-order harmonics \cite{Gauthier19}. This capability opens new avenues for designing novel refractive or diffractive optics within the solid-state medium, enabling the creation of customized harmonic beams with specific properties.

A recent breakthrough in the field has been the verification of the transfer and conservation of orbital angular momentum (OAM) from the driving laser to the high-order harmonics in both solid-state \cite{Gauthier19, Kong19} and gas-phase \cite{Gariepy14}. The generation of high-order harmonics carrying OAM was experimentally demonstrated, but the results were controversial as the law of OAM upscaling was not followed \cite{Zurch12}. These results strictly contradicted the findings from the Lewenstein model and the perturbative second harmonic generation. In \cite{Garcia13}, the authors theoretically investigated the non-perturbative process of HHG pumped by OAM beams and revealed that there is indeed an OAM buildup in different harmonic orders in accordance with the law of OAM upscaling. These findings were experimentally verified soon after~\cite{Gariepy14}. Now, it is well-established that when an intense OAM beam interacts with an atomic gas target, the OAM imprinted by the beam in the target media is conserved and that the generated harmonics obey the upscaling law of the OAM: $l_q=q\times l$ ($q$ being the harmonic order, and $l$ denotes the OAM of the driving laser beam) \cite{Paufler19}. Interestingly, the generation of extreme-ultraviolet self-torque beams, i.e., the creation of light beams with time-varying OAM, has also been demonstrated in gas-phase HHG driven by time-delayed pulses with different OAMs \cite{Laura19}. Additionally, exploring the vectorial nature of the OAM beams provides an additional degree of freedom to control this extreme nonlinear process. HHG driven by vector vortex beams (VVBs) has shown a remarkable feature: the conversion efficiency of the generated vector vortex harmonics increases with an increase in the OAM of the driving VVB \cite{Delas22}. However, to our knowledge, there is not a theoretical model that describes the interaction of an intense OAM beam with a solid medium such as a semiconductor crystal.

What sets the solid target scenario apart from the atomic case is the presence of both below and above band gap harmonics. Due to differing mechanisms, distinct power scaling laws are anticipated in each region. As vortices are sensitive to these laws, variations in vortex behavior across regions are expected, offering insights into underlying mechanisms. For instance, in atomic systems, the low photon flux limits the characterization of vortices with high topological charge \cite{Zurch12}. However, the denser nature of solid targets facilitates the characterization of such vortices. The generation and characterization of vortex beams with large topological charge hold particular significance in stimulated emission depletion microscopy (STED). Theoretically, it is possible to generate perfect vortex harmonics with higher topological charges and smaller core sizes and, consequently, improving imaging resolution. Furthermore, STED allows for the exploration of perfect optical vortices, where the intensity of the harmonic vortex remains independent of the topological charge. 

In this contribution, we explore solid-state HHG driven by an OAM beam. We theoretically investigate the non-perturbative process of HHG driven by a Laguerre–Gaussian (LG) laser beam carrying an OAM, $l=1$, in a ZnO crystal, demonstrating the conservation of OAM (i.e., the multiplicative rule for OAM), and nearly similar divergence of the emitted harmonics. To this end, we combine two well-established theories: The SBE model, which allows us to calculate the crystal response to the external field in the dipole approximation limit, and the thin-slab model, which allows us to calculate the near and far-field amplitudes of the resulting harmonic vortices.

\section{Theory of OAM laser field driven high-order harmonic generation in semiconductor materials} 

In this section, we start with a description of the vortex field and the required approximations to employ it in the semiconductor Bloch equations. Later, we introduce the SBE model followed by the thin slab model in the semiconductor material. 

\subsection{Spatiotemporal complex field}

The spatiotemporal complex field amplitude of a linearly polarized Laguerre-Gaussian (LG) beam is given by: 
\begin{eqnarray}
\label{lgbeam}
    U(r',\phi',z,t)&=&\Bigg[\frac{\omega_0}{\omega(z)}\Bigg(\frac{\sqrt2 r'}{\omega(z)}\Bigg)^{l} e^{-\Big(\frac{r'}{\omega(z)}\Big)^{2}} e^{i l \phi'} L_{P_0}^{l}\Bigg( \frac{2(r')^2}{\omega^2(z)} \Bigg) e^{i \kappa z}e^{\frac{i \kappa\; (r')^2}{2R(z)}}e^{i\varphi_{G}(z)}\Bigg]E(t),
    \end{eqnarray}
    with the temporal part $E(t)$ defined as:
    \begin{eqnarray}
    \label{et}
    E(t)&=&E_0\text{sin}^2\Bigg(\frac{\pi t}{n_c T}\Bigg)\text{sin}(\omega_L t).
\end{eqnarray}
Here, $E_{0}$, $l$, $P_0$, $\omega_{0}$, and $\varphi_G=-(2P_0+l+1)\arctan(z/z_R)$ represent the peak electric field amplitude, the OAM, the radial index, the Gaussian beam waist, and the Gouy phase of the LG beam, respectively. Here, $R(z)=z\left[1+\left(z_R/z\right)^2\right]$, and $\omega(z)=\omega_{0}\sqrt{1+\left(z/z_R\right)^2}$ represent the phase front radius, and the width of the beam at some finite propagation distance $z$, where $z_R=\kappa\omega_{0}^2/2$ is the Rayleigh length, and $\kappa=2\pi/\lambda$ is the wavenumber. In Eq.~(\ref{et}), $n_c$ is the total number of cycles in the laser field pulse, and $T=2\pi/\omega_L$ ($\omega_L$ being the central frequency of the laser) is the laser period. In our numerical simulations, we have considered a zero radial index i.e., $P_0=0$, which led to $L_{P_0}^{l}(...)=1$, to ensure that the input LG beam profile has only one radial ring. This also correspond to the experimental conditions. The used of higher values of the radial index, $P_0$, will influence the intensity distribution of the fundamental vortex beam and allow us to shape the spatial mode of the generated harmonic vortices.
It is also important to highlight that the dipole approximation is valid in our calculation. This allows us to disregard the spatial part and concentrate solely on the temporal aspect of the spatiotemporal complex field amplitude of the LG beam in the numerical solution of the SBE. This choice is justified by the characteristic lengths in our model: The lattice spacing in the ZnO semiconductor ($a_x\approx 1.4$ nm) and the electron (and hole) excursion distance, which is approximately $r_{\tau}\approx1$ nm within their respective bands. These distances are significantly smaller compared to the wavelength of the driving laser beam, $\lambda=1.55$ $\mu$m. This approximation allows us to use only the temporal part of the vortex beam in the SBE model, i.e., $E(t)$. We numerically solve the SBE equations, Eqs.~(\ref{sbes}), for a system with two bands (one conduction and one valence), considering the ZnO band structure~\cite{Vampa15}.

In semiconductor crystals, particularly in a transmission geometry setup, harmonic generation typically involves the laser beam propagating through the crystal and being focused onto the back surface. Consequently, harmonics predominantly originate in the last layers of the crystal. Harmonics generated deeper within the crystal are heavily absorbed, resulting in the primary contribution to overall harmonics occurring within the last tens of nanometers. This phenomenon allows for using a straightforward and efficient model, known as the thin-slab model (TSM) \cite{Paufler19}, as will be show in the next subsection. 

\subsection{Semiconductor Bloch equations}
Solid-HHG, particularly the interband contribution to the process, can be schematically understood in a way similar to the three-step model in atoms~\cite{amini2019symphony}. Initially, an electron residing in a valence band is pre-accelerated by the strong laser field towards the $\Gamma$-point of the Brillouin zone. Here and for ZnO, the transition probability to the conduction band reaches its maximum since at this point the energy difference between the bands is minimum \cite{ZnOProb, Navarrete}. Thus, the electron undergoes a dipole transition from the valence band to the conduction band, resulting in the creation of an electron-hole pair that is subsequently accelerated by the laser field in their respective bands. During this stage, the electron and hole experience distinct Bloch oscillations due to differences in the band structure in which their dynamics is taking place. Finally, the electron-hole pair recombines in the vicinity of the $\Gamma$-point, thereby generating high-order harmonics with a cut-off solely determined by the band gap during the time of recombination. This schematic description outlines the generation of harmonics with energies above the material's band gap. There exist however, harmonics below the energy band gap of the material. In intraband harmonic generation, the electrons (holes) are accelerated in the conduction (valence) band, resulting in the emission of harmonic radiation. It is important to highlight that the high-frequency component of the harmonic spectra results from the anharmonic electron motion \cite{TrungAnharmonic}. Additionally, for laser fluence close to the material's damage threshold, subcycle injection dynamics of electrons into the conduction band dominates the low harmonics \cite{BrunelSolids}.  

For calculating the harmonic spectra, we utilize a one-dimensional (1D)-SBE model. The fundamental premise of the SBE lies in their ability to comprehensively capture the evolution of the electron wavefunction across the reciprocal space domain. This comprehensive approach is crucial for understanding how electrons respond to the complex interplay between the crystal's periodic potential and the external laser field, particularly in scenarios where the laser field strength is sufficiently intense to induce nonlinear optical effects. The SBE can be written in terms of  the interband coherence (polarization) $p_k$ and the occupation $n_k^{e(h)}$ 
of electrons (holes) as (in a.u.):
\begin{eqnarray}
\label{sbes}
\nonumber
i\frac{\partial}{\partial t}p_k&=&\left(\varepsilon_k^e+\varepsilon_k^h-i\frac{1}{T_2}\right) p_k-(1-n_k^e-n_k^h)E(t)d_k\\
&&-iE(t)\nabla_kp_k \nonumber\\
\frac{\partial}{\partial t}n_k^e&=&-2\mathrm{Im}[ E(t)d_k p_k^{*}]-E(t)\nabla_k n_k^e\\
\frac{\partial}{\partial t}n_k^h&=&-2\mathrm{Im}[ E(t)d_k p_k^{*}]-E(t)\nabla_k n_k^h \nonumber,
\end{eqnarray}
where $\varepsilon_k^{e(h)}$ are the single particle energies of the electrons (holes), $T_2$ is the dephasing time of the polarization, $d_k$ is the ($k$-dependent) dipole transition matrix element between the valence and conduction band, and $U(r',\phi',z,t)=U(r',\phi',z)\, E(t)$, is the spatio-temporal laser driving field. From Eqs.~(\ref{sbes}) we can compute the total time-dependent interband polarization $P(t)$ and intraband current density $J(t)$ as~\cite{Trung16}: 
\begin{eqnarray}
 P(t)&=&\sum_{k}\left[ d_k\;p_k(t)+\mathrm{c.c}\right],\\   
 J(t)&=&-2\sum_{k}\left[ v_k^{e}n_k^e(t)+v_k^{h}n_k^h(t)\right],
\end{eqnarray}
where $v_k^{e(h)}$ is the group velocity of the electrons (holes) defined by $v_k^{e(h)}=\nabla_k\varepsilon_k^{e(h)}$. The total emitted spectral intensity can be calculated as:
\begin{eqnarray}
S(\omega)\propto|\omega\;P(\omega)+iJ(\omega)|^2.
\end{eqnarray}

Furthermore, for the sake of simplicity, we consider sampling the Brillouin zone only in one dimension, which corresponds to the most effective direction in the crystals. It was observed during the numerical calculations that for the 3rd harmonic, both the inter and intraband mechanisms contribute significantly to the total HHG spectrum, while for the following three harmonics, the interband polarization contributions, $P(\omega)$, dominate. For our numerical calculations, we have included both contributions for the 3rd harmonic and omitted the intraband contribution for harmonic orders 5th to 9th. It is important to clarify that scaling law calculations with the total spectral intensity yield the same results.

\subsection{Thin-slab model in semiconductors}

The TSM provides insights into how the orbital angular momentum (OAM) carried by the incident laser field is transferred to the harmonics generated in the extreme ultraviolet (XUV) regime through the non-perturbative HHG process. Here, we will particularly investigate a LG beam in the near-infrared (NIR) regime interacting with a ZnO crystal. By taking advantage of the fact that the HHG process occurs in the last layers of the crystal, we simplify the crystal representation to a thin 2D slab oriented perpendicular to the propagation direction of the incident LG beam. Combining the TSM with the 1D-SBE, we can analyze the near-field amplitude and phase profiles of various harmonic orders. Subsequently, by employing the Fraunhofer diffraction integral we can calculate the far-field amplitude and phase profiles for different harmonic orders. In our model, we position the slab precisely at the focal point of the LG beam (i.e., at $z=0$), enabling us to mitigate phase mismatch effects arising from the curved wavefront of the LG beam (focal phase) and its focusing (Gouy phase). We start by calculating the complex spatial beam profile of the fundamental LG beam in the thin slab, positioned at the beam focus i.e., at $z=0$. This profile can be written as: 
\begin{equation}
    A(r',\phi')=U(r')e^{i\Phi(\phi')}.
\end{equation}
The spatial amplitude and phase components of the fundamental LG beam, as given in Eq.~(\ref{lgbeam}), are accommodated in $U(r')$ and $\Phi(\phi')$, respectively. The next step in the model is to consider that the harmonics are emitted at the thin slab due to the nonlinear interaction of the LG beam with the atoms placed periodically. Therefore, the near-field amplitude of the $q^{\mathrm{th}}$ harmonic is calculated as:
\begin{equation}
   A^{(\text{near})}_{q}\propto \big|U(r')\big|^p e^{i q \Phi(\phi')} \label{near}.
\end{equation}
Here, $p$ is a scaling factor unique to each harmonic order and is obtained from the 1D SBE simulations. This factor describes the harmonic intensity evolution as a function of the fundamental laser field intensity. To extract the value of $p$ for different harmonics, it is necessary to calculate the harmonics amplitude scaling within a small range, as in the atomic case \cite{Garcia13}. After calculating the near-field amplitude with the help of Eq.~(\ref{near}), we use the Fraunhofer diffraction integral to compute the far-field amplitude and phase profile of the different harmonic orders. Unlike gas-phase HHG, we did not take into account the dipole phase contribution while computing the near-field complex amplitude for solids in this model. Therefore, essentially, it is the TSM with null dipole phase. The dipole phase is fundamental for a full phase matching analysis or for attosecond pulses generation. However, for our case, the intensity distribution of the vortex beams at the far-field, the phase can be safely disregarded. The far-field complex amplitude of the $q^{\mathrm{th}}$ harmonic order is thus given by: 

\begin{eqnarray}
    A_q^{(\text{far})}(\beta, \phi)&\propto& e^{iql\phi}i^{ql} \int_0^\infty r' dr' \big|U(r')\big|^pJ_{ql}\Bigg( \frac{2\pi\beta r'}{\lambda_q} \Bigg) \label{far}
\end{eqnarray}

Here, ($\beta$,$\phi$) are the far-field coordinates, representing the divergence and the azimuthal coordinate, respectively. $\lambda_{q}=\lambda/q$ denotes the wavelength of the $q^{\mathrm{th}}$ order harmonic, and $J_{ql}(...)$ represents the Bessel function of the order $ql$. In Eq.~(\ref{far}), we can observe that the phase of the $q^{\mathrm{th}}$ harmonic order scales as $\phi_{q}=q\,\phi_{f}$, where $\phi_{f}$ is the fundamental beam's helical phase and equals $l\, \phi'$. This implies that the OAM of the $q^{\mathrm{th}}$ order harmonic is equal to $q$ times the OAM of the fundamental beam, i.e., $l_{q}=q\, l$. Hence, the upscaling law of OAM  holds when HHG is driven by an LG beam in a semiconductor crystal.

Additionally, to explore the hidden spatial features of different harmonic orders, such as their intensity ring thickness and divergence profile, we calculate the far-field intensity as: 
\begin{equation}
I(\beta)\propto \big|A_q^{(\text{far})}(\beta, \phi)\big|^2,
\end{equation}
from which we can extract the far-field transverse intensity distribution for different harmonic orders:
\begin{eqnarray}
    I(\beta_x,\beta_y)&\propto&\frac{I_0^p (1)^{ql}\left(\frac{2 \pi q \sqrt{\beta_x^2+\beta_y^2}}{    {\lambda_0}}\right)^{2ql}\Gamma \left[\frac{1}{2} (pl+ql+2)\right]^2}{2^{2ql+2}\left(\frac{p}{\omega_0^2}\right)^{l p+l q+2}\Gamma[ql+1]^2}\nonumber\\
    &\times& \left\{_1F_1 \Bigg[\frac{pl+ql+2}{2};ql+1;-\frac{\pi^2 q^2 \omega_0^2 \left(\beta_x^2+\beta_y^2\right)}{p\lambda_0^2} \Bigg]\right\}^2.\label{TIP}
\end{eqnarray}

Using Eq.~(\ref{TIP}), we can calculate the intensity profile of different harmonic orders. In the following section, we will present the numerical simulation results obtained with our model, used to compute the near and far-field harmonic amplitudes, and subsequently compare them with the experimental results presented in Ref.~\cite{Gauthier19}. Notice that we have not included transverse phase matching effects in our model since we are locating the slab at the beam focus. Even though we do not include this effect in our calculations, we expect it to not change the divergence of the vortex beams, similar to what is shown for the atomic case \cite{Rego}. 

\section{Results and discussion}

\begin{figure}[h!]
\includegraphics[width=1\textwidth]{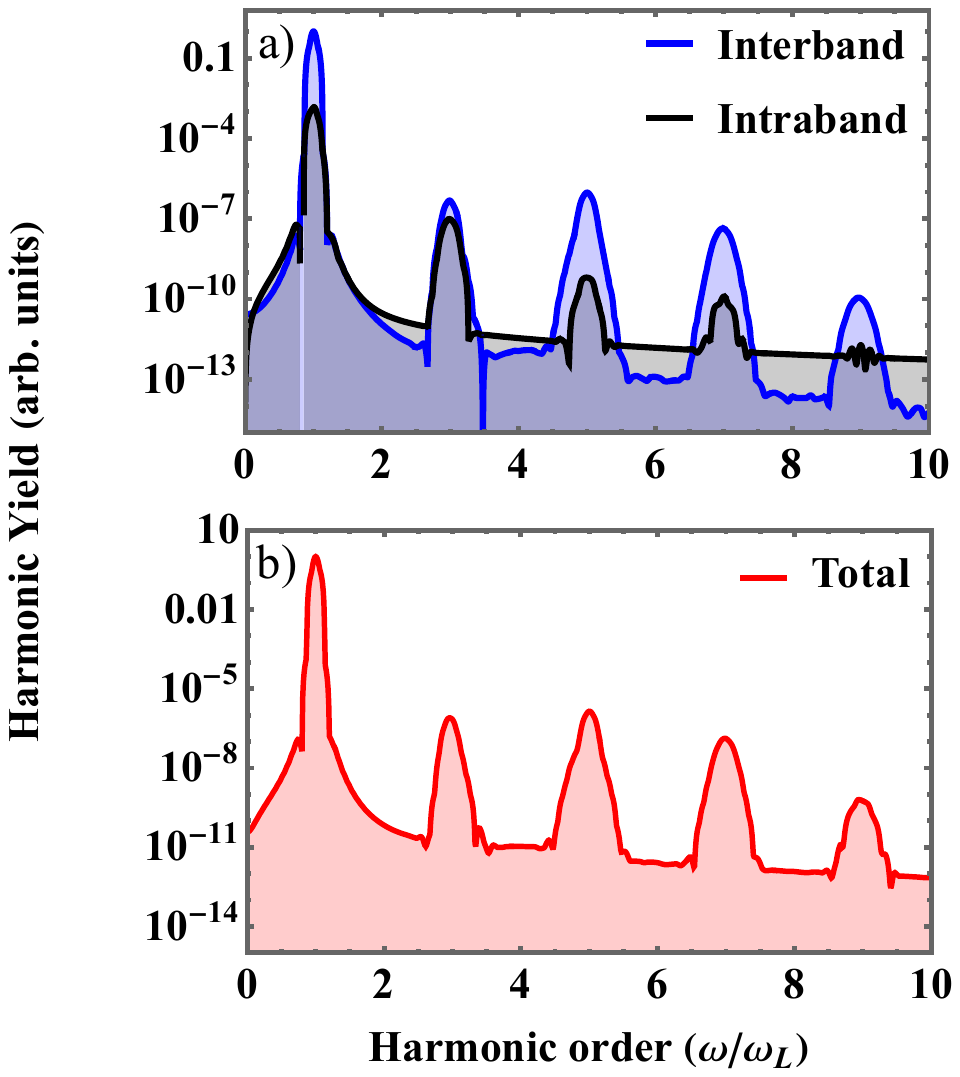}
\caption{Harmonic spectrum obtained from the 1D-SBE model for a driving laser field intensity of 2.5$\times 10^{11}$ W/cm$^2$. In a) we show a comparison between the intra and interband contributions to the total harmonic generation shown in b). The variation of the harmonic amplitudes as a function of the fundamental field amplitude is used to calculate the values of the scaling factor $p$ (see text for more details).}\label{Fig1} 
\end{figure}

We calculate twelve different harmonic spectra by solving the 1D SBE model given in Eqs.~(\ref{sbes}) to extract the scaling factor $p$. In our numerical simulation, we use a sin$^2$-shaped laser pulse with 15 total cycles and a central wavelength of $\lambda = 1.55$ $\mu$m. The total laser pulse duration is approximately 80 fs. The laser intensity varies from 2 to 2.5$\times 10^{11}$ W/cm$^2$ (corresponding to peak electric fields between and 0.0024 a.u and 0.0027 a.u.). For the calculations, we set the dephasing time to 2 fs. The small value for the dephasing time is established when considering the propagation of the laser field and harmonic field inside the bulk \cite{DephaSemi}. This makes the contribution of the short electron trajectories dominant~\cite{Vampa15}. The parameters used in our simulation are consistent with the experimental parameters used in Ref.~\cite{Gauthier19}. The calculations were performed using the orientation symmetry of the reciprocal lattice defined by $\hat{x}||\Gamma-M||$, $\hat{y}||\Gamma-K||$ and $\hat{z}||\Gamma-A||$. For this symmetry, the lattice parameters are defined by $(a_x,a_y,a_z) = (5.32, 4.61, 9.38)$ a.u., as used in Ref.~\cite{Vampa15}. For the simulations presented here, we use a linearly polarized laser electric field in $\hat{x}$ and tested the model with the $\hat{y}$ orientation, for which similar results were obtained. An example of the calculated spectrum is shown in Fig.~\ref{Fig1}.

\begin{figure}[h!]
\includegraphics[width=1\textwidth]{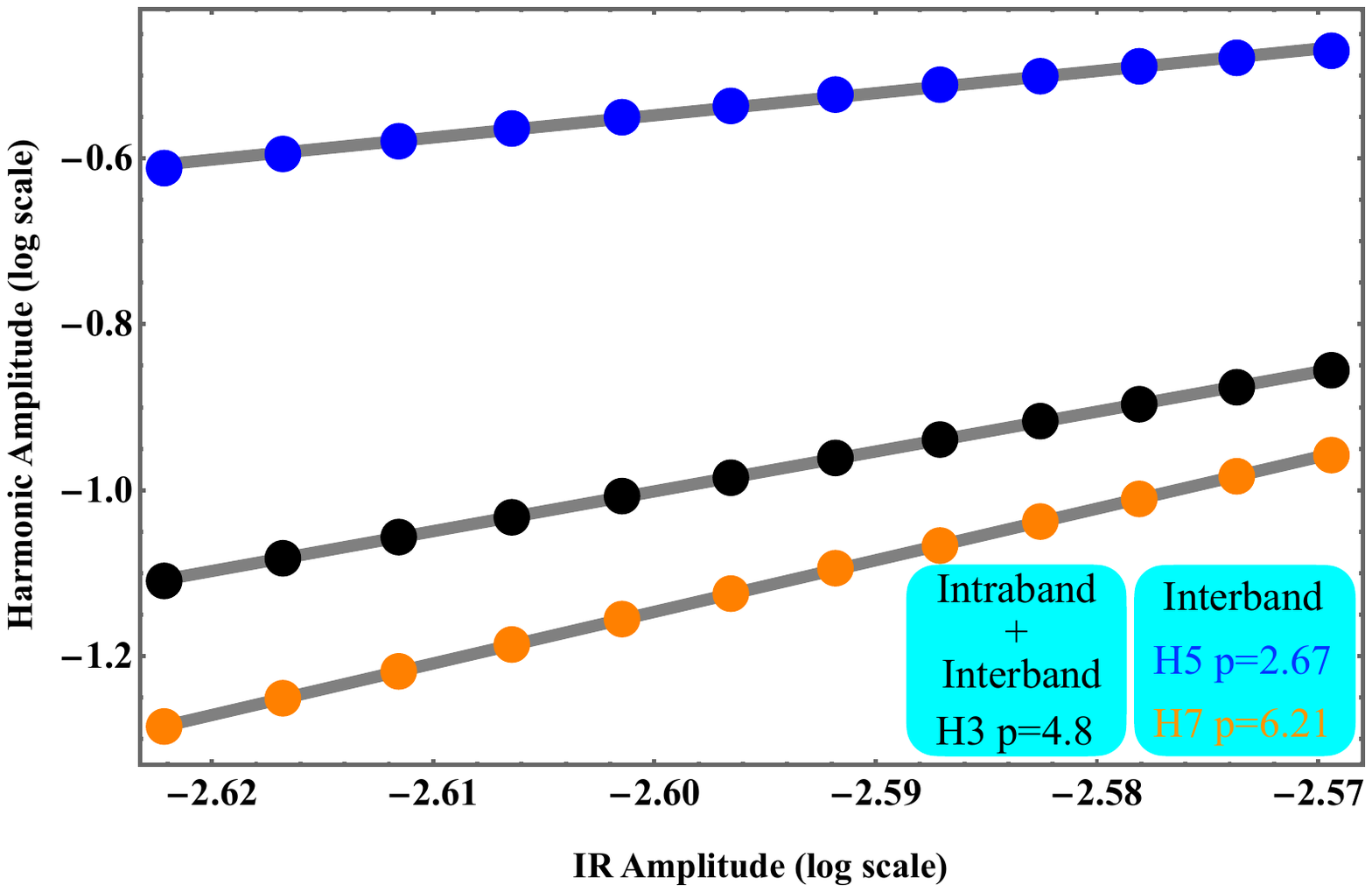}
\caption{Harmonic amplitude as a function of the fundamental laser field peak amplitude (in log scale).}\label{Fig2} 
\end{figure}

In the experimental work reported in Ref.~\cite{Gauthier19}, the analysis on the OAM transfer from the fundamental beam to the generated harmonics focused on the $3^\text{rd}$, $5^\text{th}$, and $7^\text{th}$ harmonic orders. To compare these experimental results with our theoretical calculations, we analyze the variations in the same harmonic amplitudes as a function of the fundamental field amplitude (both in log scale) as shown in Fig.~\ref{Fig2}. By fitting the resulting amplitude scaling, we extract the value of $p$ for the different harmonic orders. The resulting values of $p$ were found to be $4.8$, $2.7$, and $6.2$ for the $3^\text{rd}$, $5^\text{th}$, and $7^\text{th}$ harmonics, respectively (see Fig.~\ref{Fig2}). Notice that for the driving laser field, the photon energy is $E_{\omega_L}=0.83$ eV. Thus, $E_{\omega_3}=2.49$ eV, which is below the band gap energy of ZnO ($E_{g}=3.37$ eV). This situates the $3^\text{rd}$ harmonic in a region where neither a perturbative nor a non-perturbative scaling law is followed~\cite{Vampa15}.

\begin{figure}[h!]
\includegraphics[width=0.8\textwidth]{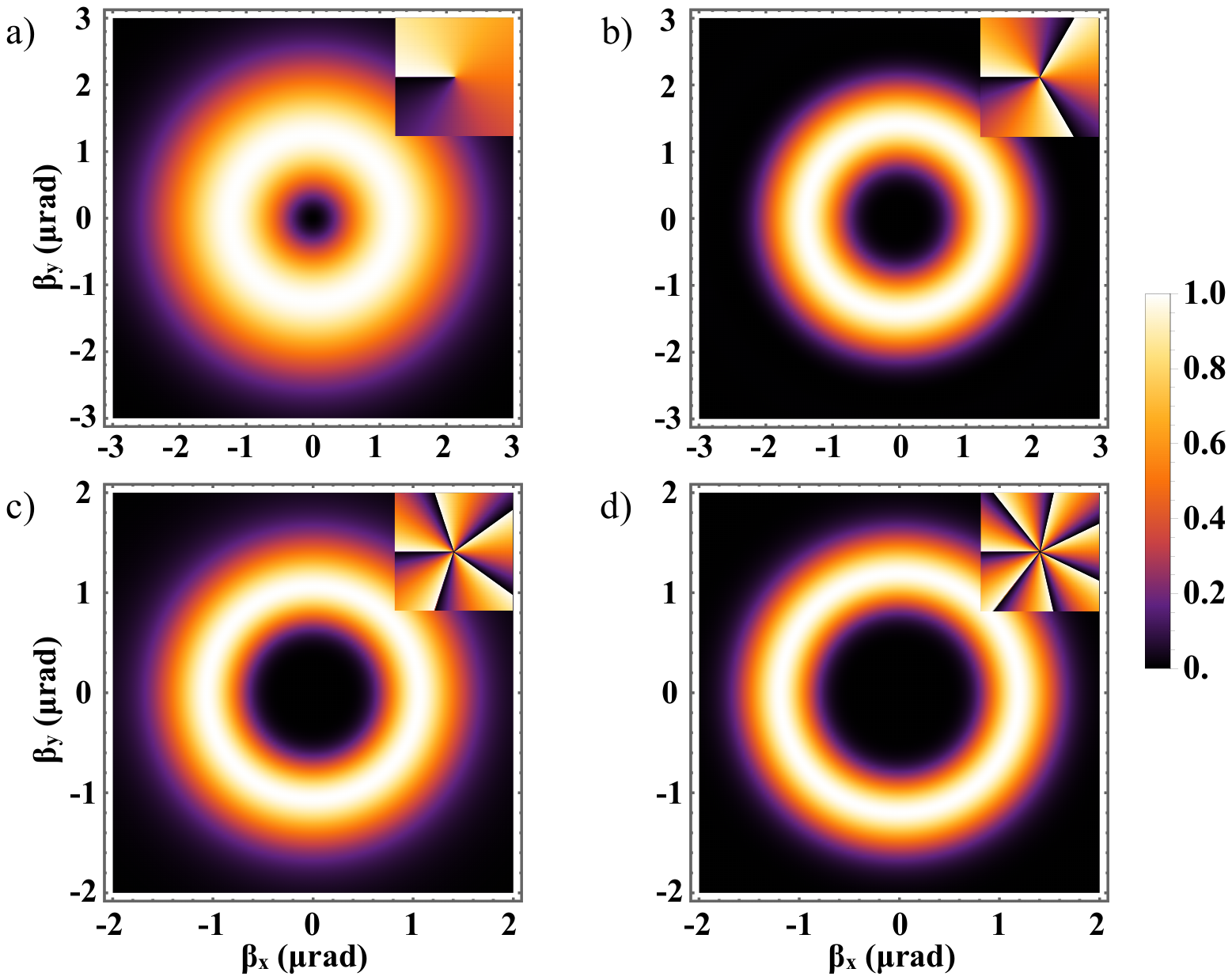}
\caption{Transverse intensity profile of the fundamental LG beam, and different harmonic orders. (a) fundamental LG beam, (b) $3^\text{rd}$ harmonic, (c) $5^\text{th}$ harmonic, and (d) $7^\text{th}$ harmonic. Their corresponding phase profiles are shown as insets. Here, the black and the white color correspond to $-\pi$, and $\pi$, respectively.}\label{Fig3}
\end{figure}

After obtaining the $p$ values for different harmonic orders, we plot their corresponding transverse intensity profiles using Eq.~(\ref{TIP}). This, in turn, provides us with some significant insights into the far-field spatial profile of the harmonic vortices. The transverse intensity profiles for the (a) fundamental LG beam, (b) $3^\text{rd}$ harmonic, (c) $5^\text{th}$ harmonic, and (d) $7^\text{th}$ harmonic are shown in Figs.~\ref{Fig3}(a)-(d), respectively. In the insets of these intensity plots, we also show the far-field phase profile for different harmonic orders, computed using Eq.~(\ref{far}). The value of the OAM for different harmonic orders can be obtained by counting the number of $2\pi$ phase shifts along the azimuthal direction, which are 3, 5, and 7 for the harmonic orders $3^\text{rd}$, $5^\text{th}$, and $7^\text{th}$, respectively. From the transverse intensity plots, two important characteristics can be observed: (i) the enhancement of the size of the central dark core with the harmonic order, and (ii) the thickness of the ring decreases with the harmonic order.

\begin{figure}[h!]
\includegraphics[width=1\textwidth]{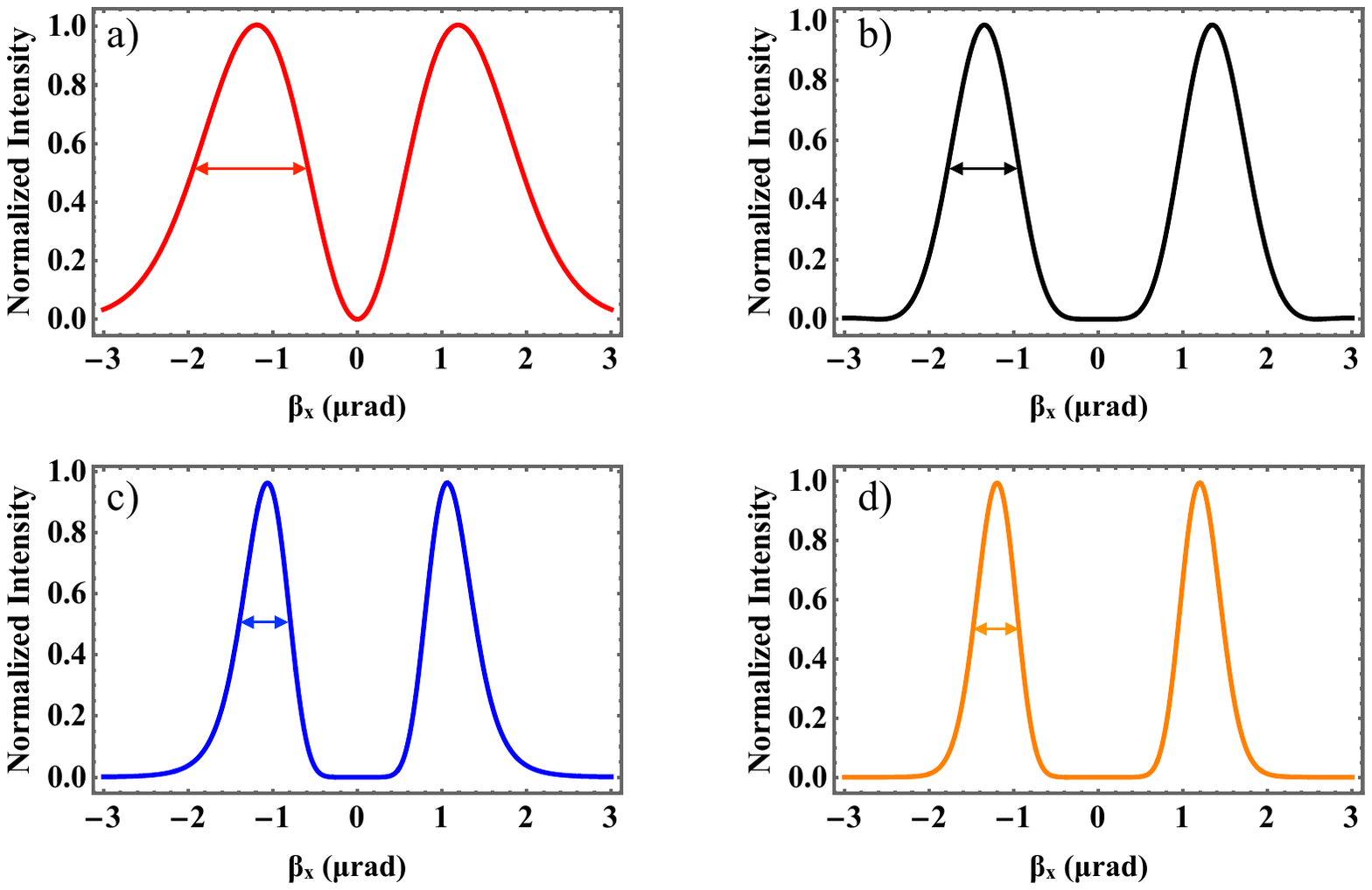}
\caption{Line profile for $I(\beta_x,0)$. (a) fundamental LG beam, (b) $3^{rd}$ harmonic, (c) $5^{th}$ harmonic, and (d) $7^{th}$ harmonic. The predicted value for the thickness of the rings at the FWHM, shown by the color horizontal arrows, is (a) 1.27 $\mu$rad, (b) 0.84 $\mu$rad, (c) 0.59 $\mu$rad, and (d) 0.54  $\mu$rad.}\label{Fig4}
\end{figure}

To quantify the vortex ring size for different harmonics, we depict the line intensity profile given by Eq.~(\ref{TIP}) for $\beta_y=0$. The profiles are shown in Fig.~\ref{Fig4}. We define the thickness of the ring for different harmonic vortices as the full width at half maximum (FWHM) of the intensity profiles. This is illustrated with the colored horizontal arrows in Fig.~\ref{Fig3}(a)-(d). The different thickness values were found to be (a) 1.27 $\mu$rad, (b) 0.84 $\mu$rad, (c) 0.59 $\mu$rad, and (d) 0.54 $\mu$rad. This demonstrates the decreasing nature of the ring thickness with the harmonic order.

 The effect of the target position relative to the laser field focus point was also investigated. By considering the thin target at different positions, we demonstrated that full phase matching effects do not alter the macroscopic response of the target. Both calculations yield the same theoretical results, which reproduce the experimental findings reasonably well, as shown in Appendix A. Hence, we can safely conclude that the transverse propagation of the laser field through the material, within the framework of the thin slab model, does not affect the macroscopic response of the solid.

It is important to highlight the necessity of experimental scaling laws to benchmark the results in this laser field intensity regime, where experimental data are available. Consequently, this will allow us to verify the value of the dephasing time, $T_2$, and determine whether this value can be extrapolated to other intensity regimes. Investigating the dependence of the scaling law on the dephasing time, $T_2$, is only possible at higher laser field intensities (see e.g.~\cite{Ghimire2011}). Therefore, studies involving vortex beams interacting with solid-state materials theoretically pave the way for establishing dephasing time values with the aid of experimental scaling laws. 

\begin{figure}[h!]
\includegraphics[width=0.6\textwidth]{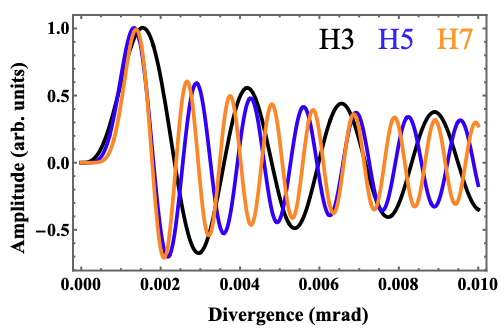}
\caption{Amplitude of the integrand in Eq.~(\ref{far}), $r_{\mathrm{max}}U^{p}(r_{\mathrm{max}},z=0)J_{ql}(2 \pi \beta r_{\mathrm{max}}/\lambda_{q})$, for $3^{\text{rd}}$ (black line), $5^{\text{th}}$ (blue line), and $7^{\text{th}}$ (orange line) harmonic orders. Here, $r_\mathrm{max}$ is the radius we obtain by maximizing $r'U^{p}(r',z=0)$. The divergence where the integrand is maximized is very similar for the harmonic orders considered in our case.}\label{Fig5}
\end{figure}

We further investigate whether the harmonics are emitted with similar divergence. It has been demonstrated both theoretically and experimentally that different harmonics are emitted with similar divergence when HHG is driven by an OAM beam in atomic gases. However, this feature has never been studied in the case of solid-HHG. For this analysis, we compute the radius at which the function $r'U^{p}(r',z=0)$ is maximum (the divergence is quantified as the radius of maximum intensity in the case of light beams carrying OAM). This maximum radius was found to be $r_{\text{max}}=\sqrt{(p l+1)/(2p)}\omega_{0}$. Then, we depict the amplitude of the integrand of Eq.~(\ref{far}), $r_{\text{max}}U^{p}(r_{\text{max}},z=0)J_{ql}(2 \pi \beta r_{\text{max}}/\lambda_{q})$ as a function of the divergence for the $3^\text{rd}$, $5^\text{th}$, and $7^\text{th}$ harmonic orders. The divergence, where the integrand is maximized, is very similar for all the harmonic orders, as shown in Fig.~\ref{Fig5}. This behavior of the solid-phase harmonics also resembles that of the gas-phase harmonics. The similar divergence for different harmonics, when HHG is driven by an OAM beam either in solid or in gas phase, can be thought of as a consequence of the OAM build-up law. In fact, this can be disentangled from the far-field complex harmonic amplitude, given by Eq.~(\ref{far}). In the TSM, the dependence on the divergence is encoded in the argument of the Bessel function, which is a part of the amplitude of the integrand. The order of this Bessel function is also related to the OAM of the $q^\text{th}$ harmonic ($ql$) generated via HHG. The argument of the Bessel function, which is directly proportional to the divergence, is also inversely proportional to the wavelength of the $q^\text{th}$ order harmonic, i.e., $\lambda_{q}$. Therefore, both the argument and the order of the Bessel function are proportional to the harmonic order. This is the reason why the position of the maximum amplitude of the integrand of Eq.~(\ref{far}) remains nearly the same for the harmonic orders considered in our case. 

It is a well-known fact that the harmonic amplitude scales differently with the fundamental field amplitude in the perturbative and non-perturbative case. In the non-perturbative case, this scaling factor lies significantly below the harmonic order. In our numerical simulation, the scaling law clearly follows this principle for the 5$^\text{th}$ and 7$^\text{th}$ harmonics, but not for the 3$^\text{rd}$ one, whose photon energy ($=2.49$ eV) lies below the band gap energy of ZnO ($=3.37$ eV). Therefore, for this harmonic, it is necessary to consider both the intraband and interband contributions to the current to calculate its corresponding $p$-value. The fact that the energy of the 3$^\text{rd}$ harmonic stands below the ZnO band gap could explain the different scaling behavior (see Fig.~\ref{Fig2}). On the other hand, even when the harmonic energy is above the band gap energy, our results suggest that a very different scaling law is observed for other individual harmonics (5$^\text{th}$ and 7$^\text{th}$ in our case), contrary to the atomic case where the scaling law is approximately the same for the harmonics located in the plateau region of the HHG spectrum~\cite{Rego17}. This is not surprising, since the dynamics of the electron are constrained by the band dispersion relation in solids, which is not the case for atomic gases, where it can be considered free, if the laser electric field is strong enough. It is expected, however, that for the plateau region, the solid HHG also have similar amplitude scaling laws. Furthermore, we tested the SBE for different dephasing times $T_2=\{1,2,5,10\}$ and 20. The resulting divergences, scaling laws and vortex sizes are not affected by changes in the dephasing time. For more details please see the Appendix B.

With our theoretical model, we observed two main characteristics of the HHG generated by the interaction of an intense LG laser field with the ZnO semiconductor: (i) The ring thickness in the transverse intensity distribution decreases as the harmonic order increases (as shown in Fig.~\ref{Fig3} for the transverse intensity distribution and in Fig.~\ref{Fig4}), and (ii) the multiplicative rule for the OAM of the harmonics, i.e., the OAM of the $q^{\mathrm{th}}$ order harmonic is given by $l_q=q\, l$ (see the insets of Fig.~\ref{Fig3}). This multiplicative rule for different harmonic orders results in the emission of harmonics with nearly similar divergence. These results are in agreement with the experimental findings of Ref.~\cite{Gauthier19}, where these features were demonstrated experimentally. The emission of different harmonics with nearly identical divergence may provide a unique route to generate attosecond helical pulses.

In a similar fashion, we calculated the divergence and intensity profile for different harmonic vortices for dephasing times ranging from 1 fs to 20 fs, as well as for various p-values. This allowed us to conclude that the ring thickness is not affected by changes in the scaling factors that may arise from the assumption of short dephasing times (see Appendix B).

\section{Conclusions and Outlook}

Our theoretical investigation into high-order harmonic generation in ZnO semiconductor, assisted by an intense Laguerre-Gaussian beam carrying orbital angular momentum (OAM) with $l=1$, reveals several key findings: (i) The scaling laws for the different harmonic orders investigated here exhibit significant deviations from reported scaling laws in the atomic case; (ii) The size of the harmonic vortices decreases as the harmonic order increases and (iii) The law of orbital angular momentum up-scaling persists, leading to the generation of harmonics with nearly identical divergence. This theoretical results align very well with the experimental findings presented in \cite{Gauthier19}. Our theoretical results, 1D SBE model in combination with the TSM, proves to be effective in describing this nonlinear interaction. 

The one-dimensional SBE is a powerful tool that, in principle, restricted to the interaction of a linearly polarized laser field with a solid target. Our model also allows for testing different crystal orientations. The SBE model for various crystal orientations predicts the same scaling laws for the harmonics investigated in this work. Additionally, the model was used to explain the creation of attosecond pulses in a ZnO crystal \cite{ParisInProgress}, showing an excellent agreement with experimental results. It is important to note that the model used in this work cannot be extended to more complex materials without employing a 2D-SBE model or in cases where the dipole approximation breaks down. Furthermore, our model successfully reproduced the experimental findings, allowing us to conclude that the dynamics of the laser field's interaction with the solid target are dominated by the vortex beam in the far-field, rather than by the dynamics of the solid itself. The solid's dynamics are responsible for the observed scaling laws.

Finally, the combination of the SBE model with the TSM allows us to accurately reproduce experimental results, largely due to the essential characteristics of the thin-slab model. In this approach, we focus on a one-dimensional semiconductor interacting with a linearly polarized laser field under the dipole approximation. In this context, the spatial dependence of the vortex field can be neglected, simplifying the electron dynamics in the material to a single-electron response within a single Brillouin zone. Consequently, the material’s full response in the far-field is reduced to the appropriate scaling of this single-electron response, which determines the far-field properties of the vortex beam. However, it is evident that without the correct implementation of the material’s response, even in the direction of the laser field polarization, extracting the scaling factor would lead to deviations from experimental findings. Our model successfully reproduces the macroscopic experimental observations based on the microscopic response of the material to a time-varying laser field. If the dipole approximation were to break down, the electron dynamics in the SBE model would need to incorporate the spatial aspects of the vortex beam. This is particularly crucial for two-dimensional materials, where symmetry plays a significant role, and nonlinear laser polarization requires a complete solution that includes the spatial component of the vortex beam.

Our approach to solid-state HHG driven by shaped light presents opportunities to explore selection rules considering various degrees of freedom, such as OAM, spin angular momentum (SAM), the symmetry of the generating medium, and the laser field. Furthermore, future investigations may involve applying this theoretical approach to analyze the interaction of solid-state media with different spatially structured beams, such as Bessel-Gauss (BG) and perfect optical vortex (POV) beams. This offers avenues for deeper exploration into the nonlinear dynamics of solid-state HHG. Additionally, a full phase-matching analysis is required when comparing harmonics from the bulk with those from the surface, for which our model provides a convenient solution. The model can also be applied to investigate transmission and reflection HHG in solid targets.

\section*{Acknowledgements}
We acknowledge financial support from the Guangdong Province Science and Technology Major Project (Future functional materials under extreme conditions - 2021B0301030005) and the Guangdong Natural Science Foundation (General Program project No. 2023A1515010871).

\section*{Appendix A: Thin slab model for $z\neq0$}

The calculations presented in Eq.~(\ref{TIP}) and the conclusions drawn in the manuscript are based on the assumption that the target is positioned at the focus of the driving laser beam, allowing propagation effects to be neglected. However, it is crucial to calculate, within the framework of the thin slab model (TSM), the full propagation effects and demonstrate that they do not alter the macroscopic response of the material to the strong laser field. With this in mind, we calculate the macroscopic response within the TSM model for $z\neq0$.
% Even withing the framework of the TSM, the consideration of the full propagation effects do not change the general trend of the macroscopic response of the material to the strong laser field. To observe this behaviour, the target is placed at some finite values of $z$ (apart from being at the laser focus as discussed earlier in the main-text).  

\begin{figure}[h!]
\includegraphics[width=1\textwidth,angle =-90]{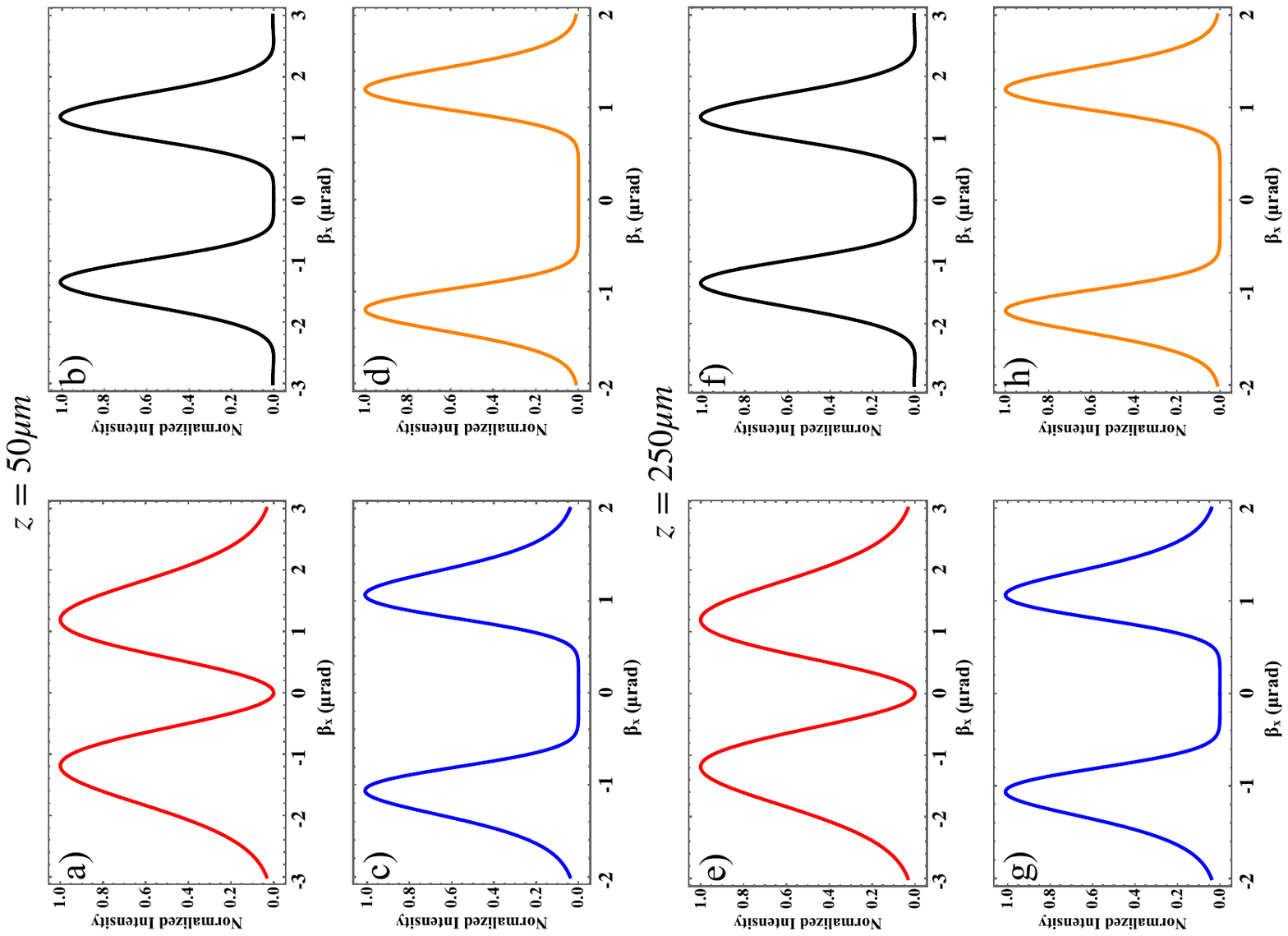}
\caption{Vortex size for different values of $z$. The values for the FWHM for the harmonic vortices are a) H1=1.40 $\mu$rad, b) H3 = 0.80 $\mu$rad, c) H5 = 0.63 $\mu$rad, d) H7 = 0.54 $\mu$rad for z = 50 $\mu$rad. Similar values were found for panels e) to h). }\label{Fig6}
\end{figure} 

Calculating the far-field intensity profiles for different harmonic orders, when the target is located at various values of $z$, requires integrating the full expression of the driving laser field presented in Eq.~(\ref{lgbeam}). This process leads to the following expression for the far-field intensity (note that the different phase contributions cancel out when evaluating the intensity)

\begin{eqnarray}
    I(\beta_x,\beta_y)&\propto&\Bigg|\frac{2\pi E_0^p i^{ql}\left(\frac{\omega_0}{\omega(z)}\right)^p\left(\frac{\sqrt{2}}{\omega(z)}\right)^{l p}\left(\frac{2 \pi q \sqrt{\beta_x^2+\beta_y^2}}{    {\lambda_0}}\right)^{ql}\Gamma \left[\frac{1}{2} (pl+ql+2)\right]}{2^{ql+1}\left(\frac{p}{\omega^2(z)}-\frac{iqk}{2R(z)}\right)^{\frac{l p+l q+2}{2}}\Gamma[ql+1]}\nonumber\\
    &\times& \left\{_1F_1 \Bigg[\frac{pl+ql+2}{2};ql+1;-\frac{\left(2\pi\frac{q}{\lambda_0} \sqrt{\beta_x^2+\beta_y^2}\right)^2}{4\left(\frac{p}{\omega(z)^2}-\frac{iqk}{2R(z)}\right)} \Bigg]\right\}\Bigg|^2,\label{TSMz}
\end{eqnarray}
with $ R(z) = z \left( 1 + \frac{z_R^2}{z^2} \right) $ as the wavefront radius of curvature, $\omega(z) = \omega_0 \sqrt{1 + \frac{z^2}{z_R^2}}$ as the beam width, $z_R = \frac{\pi \omega_0^2}{\lambda_0}$ as the Rayleigh length, and the wave number $k = \frac{2\pi}{\lambda_0}$, we computed the harmonic intensity profiles for different harmonic orders using Eq.~(\ref{TSMz}) for various positions of the slab (representing the target in the TSM) relative to the laser focus. In Fig.~\ref{Fig6}, we present the results for two different slab locations, i.e., at $z = 50$ $\mu$m and $z = 250$  $\mu$m. It is clear from the figure that the inclusion of the full propagation effects does not alter the observed results presented in Fig.~\ref{Fig4}. A comparison between Fig.~\ref{Fig4} and Fig.~\ref{Fig6} reveals that the sizes of the harmonic vortices are not exactly the same; however, the ring thickness decreases and the vortex core size increases as the harmonic order increases, consistent with experimental measurements. This is evident from the FWHM values extracted from the line intensity plots in Fig.~\ref{Fig6}, which represent the ring thickness. For different harmonics at  $z = 50$ $\mu$m, we found H1 = 1.40  $\mu$rad, H3 = 0.80 $\mu$rad, H5 = 0.63 $\mu$rad, and H7 = 0.54 $\mu$rad, clearly agreeing with the experimental conclusions drawn in \cite{Gauthier19}.

\begin{figure}[h!]
\includegraphics[width=1\textwidth,angle =0]{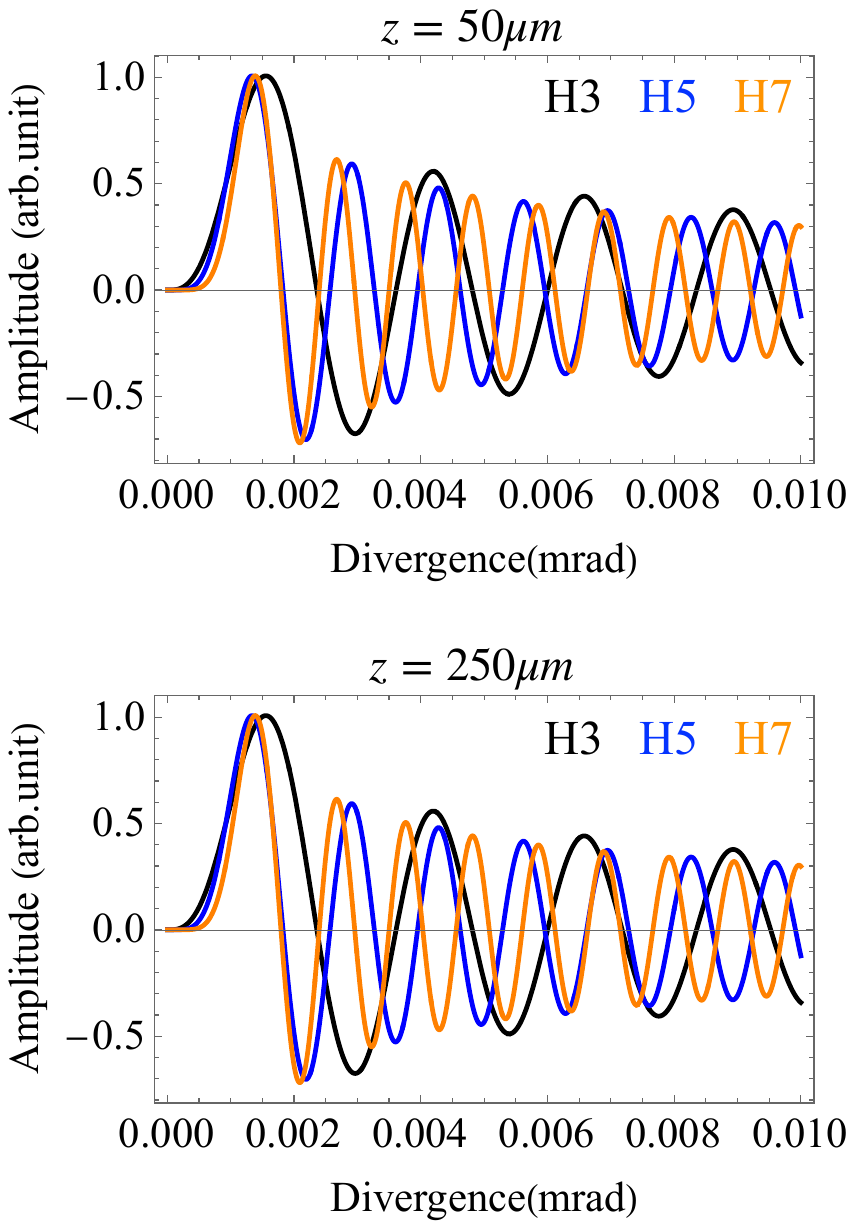}
\caption{Amplitude of the integrand in Eq.~(\ref{far}) (after including the $z$-dependence of the laser field), $r_{\mathrm{max}}U^{p}(r_{\mathrm{max}},z)J_{ql}(2 \pi \beta r_{\mathrm{max}}/\lambda_{q})$, for the $3^{\text{rd}}$ (black line), the $5^{\text{th}}$ (blue line), and the $7^{\text{th}}$ (orange line) harmonic orders at $z=50$ $\mu$m (top panel) and $z=250$ $\mu$m (bottom panel). Here, $r_\mathrm{max}$ is the radius we obtain by maximizing $r'U^{p}(r',z)$. The divergence where the integrand is maximized is very similar for the harmonic orders considered in our case. }\label{Fig8}
\end{figure}

We also investigated the effect of positioning the target at different values of $z$ on the divergence profile of the harmonic orders. The results for placing the target at $z = 50$  $\mu$m and $z = 250$ $\mu$m are presented in Fig.~\ref{Fig8}. It can be seen that the 3rd, 5th, and 7th harmonics were emitted with nearly identical divergence in both cases. Therefore, even with the inclusion of full propagation effects, the divergence profile does not change significantly compared to the case at $z = 0$. Hence, the law of OAM upscaling is still followed when the target is located at finite values of $z$.

\clearpage

\section*{Appendix B: Effect of the dephasing time on the harmonic vortices}

An important question arises regarding the accuracy of the results for different dephasing times, $T_2$. To address this, we calculated the scaling laws (p-values) for the 3rd to 7th harmonics using the 1D-SBE model across various dephasing time values.

\begin{figure}[h!]
\includegraphics[width=1\textwidth,angle =-90]{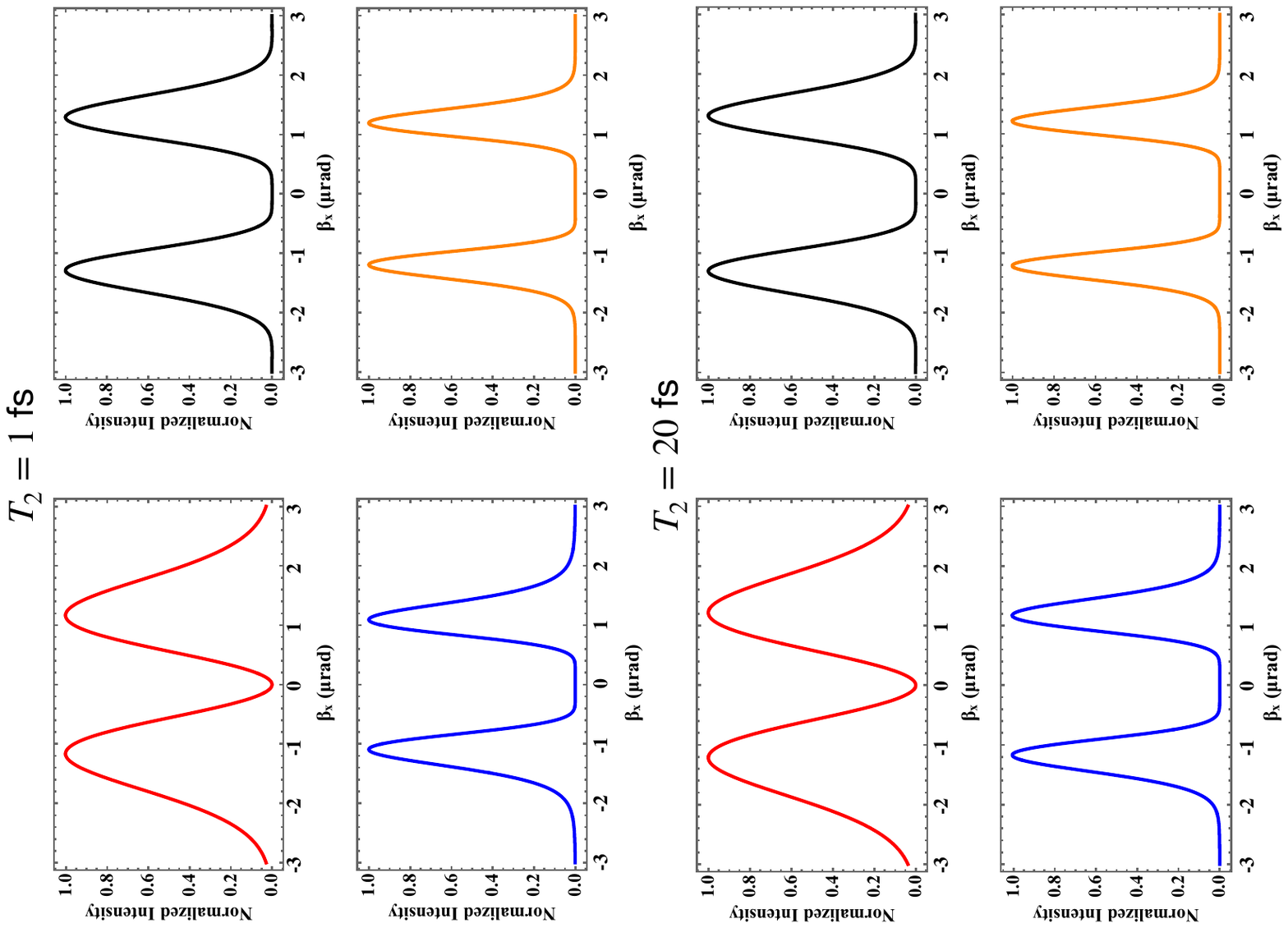}
\caption{Normalized intensity profiles as a function of the divergence of the fundamental (red), H3 (black), H5 (blue), and H7(orange) for two different dephasing times. }\label{Fig7}
\end{figure} 
The results for two different $T_2$ values are presented in Fig.\ref{Fig7}. These results clearly demonstrate that, within the framework of the SBE model, the dephasing time does not affect the scaling laws for the parameters used in this calculation. We also calculated the vortex behavior for $T_2=5$ and 10 fs (not shown here), and for these dephasing times, the results are similar to those presented in Fig.~\ref{Fig7}.

\bibliography{literatur}

%\begin{comment}

%\end{comment}

\end{document}